\documentclass{article}
\usepackage{varioref}
\usepackage{amssymb}
\usepackage{color}

\begin{document}

%&latex

%&latex
\title{\bf  Quantum time dilation in the near-horizon region of a black hole}
\author{H. Hadi\thanks{%
email: hamedhadi1388@gmail.com},~ K. Atazadeh\thanks{%
email: atazadeh@azaruniv.ac.ir},~ F. Darabi\thanks{%
email:f.darabi@azaruniv.ac.ir}
\\{\small Department of Physics, Azarbaijan Shahid Madani University, Tabriz, 53714-161 Iran}}
\date{\today} \maketitle
\begin{abstract}
In this work, we obtain a relation for average quantum time dilation between two clocks $A$ and $B$ in the near-horizon region of a black hole supported by the Rindler metric and conformal tortoise coordinate. It is indicated that this relation is identified with time dilation in classical and flat background limits.
\end{abstract}
\section{Introduction}
Time indicates itself as a classical parameter in the Schrödinger equation. This parametric feature of time is not in harmony with other dynamical features of the quantum system such as position and momentum. One can define this interpretation of time as a classical clock in the laboratory which seems to be acceptable for all practical purposes. However, quantum mechanics with its nontrivial features subjected to practical and experimental phenomena demands a fully quantum description of time. Some afforests have been tried to pose proposal for this problem \cite{1,2,3,4,5,6,7,8,9,10,11}. However, there are some controversial issues surrounding these proposals which one can refer to \cite{12, 13, 14, 15, 16}. Page and Wootters' mechanism (PaW) is one of these proposals considering time as a quantum degree of freedom \cite{17,18,19,20}. Therefore in this paradigm, all of the variables are defined by operators and there is no disconformity of time with other dynamical features of the system as a whole one of the variables is assumed to be a clock system which evaluates the dynamical evolution of the rest of system.

In this work, by restricting the model proposed here to the PaW mechanism, we consider the time dilation in general relativity. Time plays a fully different role in general relativity than Newtonian physics when asserting that time is the quantity indicated by clock \cite{clock}. This definition is completely different than Newton's abstract and absolute time which passes simultaneously for all observers in absolute space. This paradigm shift from an abstract view of time to measuring time by clocks ticking with different lapses for different observers is a good starting point for scrutinizing time in general relativity by inserting quantum effects. This goal is done by the PaW approach which starts from a timeless picture and derives the Schrödinger for subsystems.

The PaW approach has been considered in \cite{main} for flat space-time to obtain the time dilation between two clocks. The generalization to curve space-time is one of the purposes of this work. However, we have restricted the curve space-time to the near horizon region of the black hole and it is important to note the generalization.

However, note that there are some criticisms of the PaW mechanism \cite{7}. The criticisms are that in PaW formalism the PaW conditional probability: (I) leads to wrong localization probabilities for a relativistic particle, (II) violates the Hamiltonian constraint, and (III) yields incorrect transition probabilities. \\The cases (II) and (III) are discussed and solved in \cite{smith}. The criticism (I) has been resolved in \cite{smith2021}. Since we have restricted our model to \cite{main}, it is free of all criticisms of reference \cite{7}. On the other hand, there are also some other criticisms of the PaW approach which one can refer to \cite{quantumtime1,quantumtime2} for more discussions and solutions. In addition, the PaW approach and some criticisms of it also have been considered in the near horizon region of a black hole in \cite{farhad1,farhad2}. The other approach which overcomes the objections posed in \cite{7} can be found in \cite{conditional}. 

The organization of the paper is as follows. The constraint description of relativistic particles
with internal degrees of freedom is reviewed in section 2. Page Wootters formulation of $N$ particles in curved space-time is the topic of section 3 where the PaW mechanism in the near-horizon region of a black hole is supported by Rindler background and tortoise coordinate. In sections 4 and 5 quantum time dilation in Rindler background and by tortoise coordinate are considered, respectively. In the end, we have a conclusion in section 6.

\section{Constraint description of relativistic particles with internal degrees of freedom}
In this section we review the Hamiltonian constraint formulation of $N$ relativistic particles with an internal degree of freedom \cite{hamiltonian}. However, definitions that are necessary for the PaW approach in curved space-time are also explained.

Suppose $N$ free relativistic particles each carry a set of the internal degree of freedom in the general background. For our goal, these particles are moving in curved space-time background with metric $g_{\mu\nu}$. The configuration variables and its conjugate momentum are $q_{n}$ and $p_{q_{n}}(n=1,...,N)$, respectively. The action of this system is given by
\begin{equation}
	S=\sum_{n}\int d\tau_{n}L_{n}(\tau_{n}),
\end{equation}
where the Lagrangian is given by
\begin{equation}
	L_{n}(\tau_{n})=-m_{n}c^{2}+p_{q_{n}}\frac{dq_{n}}{d\tau_{n}}-H_{n}^{clock}.
\end{equation}
This is the Lagrangian of the nth particle with proper time $\tau_{n}$ and rest mass $m_{n}$. The local Hamiltonian indicating internal degrees of freedom is $H_{n}^{clock}=H_{n}^{clock}(q_{n},p_{n})$  tracking the nth particle's proper time. The spacetime position of the nth particle's center-of-mass is $x_{n}^{\mu}$ which is relative to an inertial observer. The differential proper time $d\tau_{n}$ which is along the nth particle's world line  is given by
\begin{equation}
	d\tau_{n}=(-g_{\mu\nu}\dot{x}^{\mu}_{n}\dot{x}_{n}^{\nu}/c^{2})^{1/2}dt_{n}=\sqrt{-\dot{x}_{n}^{2}}dt_{n},
\end{equation}
here the dot denotes differentiation to $t_{n}$ and shorthand form of $\dot{x}^{2}_{n}=g_{\mu \nu}\dot{x}_{n}^{\mu}\dot{x}_{n}^{\nu}$ is used.
The action takes the following form
\begin{equation}\label{action}
	S=\sum_{n}\int dt_{n} \sqrt{-\dot{x}^{2}_{n}}L_{n}(t_{n}).
\end{equation}
The action $(\ref{action})$ is invariant under world line parameter $t_{n}$. This allows the action to be parameterized in terms of a single parameter $t$. The action is $S=\int dt L(t)$ with Lagrangian
\begin{equation}\label{leg}
	L(t)=\sum_{n}\sqrt{-\dot{x}_{n}^{2}}(-m_{n}c^{2}+\frac{p_{q_{n}}\dot{q}_{n}}{\sqrt{-\dot{x}_{n}^{2}}}-H_{n}^{clock}).
\end{equation}
The temporal, spatial, and internal degrees of freedom are dynamical variables on equal footing in this Lagrangian. An inertial observer describes these variables in extended phase space. By applying Legendre transform to equation $(\ref{leg})$ one can obtain the Hamiltonian
\begin{equation}
	H=\sum_{n}[g_{\mu \nu}p^{\mu}_{n}\dot{x}^{\nu}_{n}+p_{q_{n}}\dot{q}_{n}]-L(t),
\end{equation}
\begin{equation}
	H=\sum_{n}[g_{\mu \nu}p^{\mu}_{n}\dot{x}^{\nu}_{n}+\sqrt{-\dot{x}^{2}_{n}}(m_{n}c^{2}+H_{n}^{clock})],
\end{equation}

\begin{equation}
	H_{n}=\frac{g_{\mu \nu}\dot{x}^{\mu}_{n}\dot{x}^{\nu}_{n}}{\sqrt{-\dot{x}^{2}_{n}}}M_{n}+\sqrt{-\dot{x}^{2}_{n}}(m_{n}c^{2}+H_{n}^{clock}),
\end{equation}
\begin{equation}
	H_{n}=\frac{\dot{x}^{2}_{n}}{\sqrt{-\dot{x}^{2}_{n}}}[(m_{n}c^{2}+H_{n}^{clock})-(m_{n}c^{2}+H_{n}^{clock})] \approx 0,
\end{equation}
where we have used $M_{n}= (m_{n}c^{2}+H_{n}^{clock}/c^{2})$ and ``$\approx$" means that the Hamiltonian vanishes as a constraint. Also, $p_{n}^{\mu}$ is the momentum conjugate for the nth particle's spacetime position $x_{n}^{\mu}$ given by
\begin{equation}
	p^{\mu}_{n}=g^{\mu \nu}\frac{\partial L}{\partial \dot{x}^{\mu}_{n}}=\frac{\dot{x}^{\mu}_{n}}{\sqrt{-\dot{x}^{2}_{n}}}M_{n}.
\end{equation}
The constraints can be expressed as follows
\begin{equation}\label{c12}
	C_{H_{n}}=g_{\mu \nu}p_{n}^{\mu}p_{n}^{\nu}c^{2}+M_{n}^{2}c^{4} \approx 0.
\end{equation}
This is a collection of primary first-class constraints \cite{Dirac}. One can factorize each of these constraints as $C_{H_{n}}=C_{n}^{+}C_{n}^{-}$ which is given by
\begin{equation}\label{constaint}
	C_{n}^{+}=\sqrt{g_{00}}(p_{n})_{0}+h_{n} ~~~and~~~C_{n}^{-}=\sqrt{g_{00}}(p_{n})_{0}-h_{n},
\end{equation}
where $h_{n}$ is
\begin{equation}\label{h}
	h_{n}=\sqrt{g_{ij}p_{n}^{i}p_{n}^{j}+M_{n}^{2}c^{4}}.
\end{equation}
Note that in the whole of this work, we have restricted to solutions in which the time-space components of the metric vanish $g_{oi}=0$. Maybe one can consider the cross terms of the metric too for example for the Kerr black hole solution to recognize the influence of this term in the results. However for simplicity, we start from the near horizon metric of Schwarzschild black hole, Rindler space, then we generalize it to tortoise coordinate which is a more convenient metric to work in the near-horizon region of a black hole, in the next sections. In the rest of the paper when we use the term black hole we mean Schwarzschild black hole.

\section{Page-Wootters formulation of $N$ particles in curved space-time}
In this section we consider the PaW mechanism \cite{paw} respecting $N$ relativistic particles in curved space-time. However, to consider the PaW mechanism, the background is restricted to the near horizon region of a black hole to indicate an explicit formula for the Schrödinger equation. To do so, the physical normalized state of $N$ particles is defined by $|\Psi\rangle\rangle$. Then, by enlarging the Hilbert space $\mathcal{H}_{s}$ under consideration of clock and system we have
\begin{equation}
	\mathcal{H}=\mathcal{H}_{T}\otimes\mathcal{H}_{S},
\end{equation}
where the system under consideration $\mathcal{H}_{S}$ consists of Hilbert space of center-of-mass $\mathcal{H}_{n}^{cm}$ and internal degrees of freedom (clock) $\mathcal{H}_{n}^{clock}$ is given by
\begin{equation}
	\mathcal{H}_{S}=\otimes_{n}\mathcal{H}_{n}^{cm}\otimes \mathcal{H}_{n}^{clock}.
\end{equation}
The Hilbert space $\mathcal{H}_{T}$ is for an ancillary system $T$ which is a "Clock" \footnote{Note that the clock system is indicated by operator $T$ in Hilbert space $\mathcal{H}_{T}$ and we use "C" and "Clock" with capital "C" as a representation of the ancillary clock system. However, the up-script "clock" in Hilbert space $\mathcal{H}^{clock}$ and Hamiltonian $H_{n}^{clock}$ which do not start with capital "C",  refer to internal degrees of the system which is called as the clock.} system and can be isomorphic to the Hilbert space of particle in a line. One choice other interpretations too \cite{19,rovelli}. For $N$ relativistic particles as a whole in the flat background or this model in curved space-time, the ancillary Hilbert space $\mathcal{H}_{T}$ consists of eigenstates $|t_{n}\rangle$ which temporal degree of freedom of system associated with eigenvalue $t \in R$ as follows
\begin{equation}
	T|t_{n}\rangle=t|t_{n}\rangle.
\end{equation}
The Hilbert space $\mathcal{H}_{T}$ is equipped by canonical coordinate $T$ and $P_{T}$ which satisfy $[T, P_{T}]=i$ which represent in this case time and energy indicator of an evolving system.  The constant operator of the model can be written by
\begin{equation}\label{hcrossh}
	H=\hslash P_{T}\otimes I_{S}+I_{S}\otimes H_{S},
\end{equation}
where $I_{T}$ and $I_{S}$ are the identity operator on $\mathcal{H}_{T}$ and $\mathcal{H}_{S}$, respectively. $H$ is selfadjoint and has a continuous spectrum including all possible real values and generalized eigenvalues. By applying $H$ over the physical state $|\Psi\rangle\rangle$ one can reach to null eigenvalue as follows
\begin{equation}\label{wheeler}
	H|\Psi\rangle\rangle=0.
\end{equation}
Equation $(\ref{wheeler})$ can be interpreted as the Wheeler-DeWitt equation which is a constraint forcing the physical vectors to be eigenstates of Total Hamiltonian $H$ with null eigenvalue \cite{13,de30}. By applying $I_{S}$ and projector $\Pi_{t}=|t\rangle\langle t|$ over physical state $|\Psi \rangle \rangle$ we have
\begin{equation}\label{project}
	\Pi_{t}\otimes I_{S}|\Psi\rangle \rangle = |t \rangle |\psi_{s}(t)\rangle.
\end{equation}
Note that $\Pi_{t}$ is a projector operator onto the subspace of $\mathcal{H}$ in which the temporal degree of freedom of each particle is in a definite temporal state $|t\rangle=\otimes_{n}|t_{n}\rangle$ associated with eigenvalue $t$. By using equation $(\ref{project})$ the conditional state is given by
\begin{equation}\label{conditional}
	|\psi_{s}(t)\rangle=\langle t| \otimes I_{S}|\Psi\rangle \rangle \in \mathcal{H}_{S},
\end{equation}
here the state $|\psi_{s}(t)\rangle$ indicates the states of internal degrees of freedom and system of center-of-mass of $N$ relativistic particles which are on their temporal degree of freedom in the state $|t_{n}\rangle$. By the following equation we demand the normalization condition over the state of the system for all $t \in R$
\begin{equation}
\langle \langle \Psi| \Psi \rangle \rangle _{PaW}=\langle \langle \Psi | \Pi_{t} \otimes I_	{S}|\Psi\rangle\rangle =\langle\psi_{s}(t) | \psi_{s}(t)\rangle=1.
\end{equation}
The protective valued measures (PVM) $\{\Pi_{t} , \forall  t \in R \}$ are on Hilbert space $\mathcal{H}_{T}=\otimes_{n} \mathcal{H}_{n}^{C}$. There are the following features of this space
\begin{equation}
	\langle t|\acute{t}\rangle =\delta (t-\acute{t})  ~~~,~~~\int dt \Pi_{t}= I^{C},
\end{equation}
where $I^{C}$ is the identity in Hilbert space $\mathcal{H}_{T}$. According to conditional state $(\ref{conditional})$ one can recognize that the physical state $|\Psi\rangle\rangle$ is entangled relative to $\mathcal{H}_{n}^{cm}\otimes \mathcal{H}_{n}^{clock}$ as follows
\begin{equation}
	|\Psi\rangle\rangle=\left(\int dt\Pi_{t} \otimes I_{S}\right)|\Psi\rangle\rangle= \int dt |t\rangle|\psi_{s}(t)\rangle.
\end{equation}
In a more accurate interpretation, the state $\int |t\rangle |\psi_{s}(t)\rangle$ is partitioning of kinematical Hilbert space $\mathcal{H}_{T}\otimes \mathcal{H}_{S}$ \cite{Dirac, d1,d2,d3,d4,d5} which is a concept from Dirac's constraint quantization algorithm. However, the entanglement mentioned above, is not physical which means that does not gauge invariant \cite{smith}.

To derive Wheeler-DeWitt type equation $(\ref{wheeler})$ in curved space-time which in our case is the near horizon region background we consider it in Rindler space and its conformal metric in the following subsections

\subsection{PaW mechanism in Rindler space}

 The metric of the near horizon of a black hole behaves like a Rindler space metric and is given by
\begin{equation}\label{metric}
ds^{2}=-\rho^{2}d \tau^{2}+d\rho^{2},
\end{equation}
where $\rho$ is the proper distance from the horizon of a black hole recognizing by a static Rindler observer standing in the near horizon region. Therefore the location of the horizon is $\rho=0$. To expand the PaW mechanism in the  Rindler background we consider physical states that satisfy
\begin{equation}\label{C}
	C_{n}^{+}|\Psi\rangle\rangle=(\sqrt{g_{00}}(p_{n})_{0}+h_{n})|\Psi\rangle\rangle=0,
\end{equation}
where we used equation $(\ref{constaint})$ and $h_{n}$ is operator given by equation $(\ref{h})$. Note that we will indicate $h_{n}$ by $h_{\rho n}$ and $h_{zn}$ for Rindler and tortoise backgrounds, respectively, in the rest of the paper. The equation $(\ref{C})$ is true for all $n$. In this case, $g_{00}$ and $g_{ij}$ are components of Rindler metric. To clarify the case we start by applying metric $(\ref{metric})$ to equation $(\ref{c12})$ which leads to
\begin{equation}\label{ch}
C_{H_{n}}=g_{00}p^{0}_{n}p^{0}_{n}c^{2}+M_{n}^{2}c^{4}+g_{11}p^{1}_{n}p^{1}_{n}c^{2} \approx 0
\end{equation}
Then by inserting the metric $(\ref{metric})$ into above equation we have
\begin{equation}
C_{H_{n}}=-\rho^{2}(p_{n}^{0})^{2}c^{2}+(p_{n}^{1})^{2}c^{2}+(m_{n}+\frac{H_{n}^{clock}}{c^{2}})^{2}c^{4} \approx 0.
\end{equation}
Using above equation one can obtain
\begin{equation}
\sqrt{\rho^{2}(p_{n}^{0})^{2}c^{2}}=\sqrt{(p_{n}^{1})^{2}c^{2}+(m_{n}+\frac{H_{n}^{clock}}{c^{2}})^{2}c^{4}}=h_{\rho n}.
\end{equation}
The following definitions are used for convenience
\begin{equation}
p_{n}^{0} c= h_{\rho n}/\rho = \tilde{h}_{\rho n}.
\end{equation}
Then for $C_{n}^{+}$ we have
\begin{equation}\label{cplus}
	C_{n}^{+}= (p_{n})_{0}+\tilde{h}_{\rho n}.
\end{equation}
Now by expanding $h_{\rho n}$ and ignoring quadratic interaction terms, it is given that
\begin{equation}
h_{\rho n}=m_{n}c^{2} + H_{n}^{cm}+ H_{n}^{clock}-\frac{1}{m_{n}c^{2}}[H_{n}^{cm}\otimes H_{n}^{clock} + \frac{1}{2}(H_{n}^{cm})^{2}],
\end{equation}
which leads to
\begin{equation}\label{hro}
\tilde{h}_{\rho n}=\frac{1}{\rho}\left(m_{n}c^{2} + H_{n}^{cm}+ H_{n}^{clock}-\frac{1}{m_{n}c^{2}}[H_{n}^{cm}\otimes H_{n}^{clock} + \frac{1}{2}(H_{n}^{cm})^{2}]\right).
\end{equation}
In the following, we derive the Schrödinger equation from the PaW approach in the Rindler background. The conditional state $|\psi_{s}(t)\rangle$  will obey Schrödinger equation in the parameter $t$. Note that $x_{n}^{0}$ is applied as a time operator and does the same as operator $T$ in the previous section. This choice is only because of appropriate representation when one works in the Schrödinger framework. Recall that the operators $p_{n}^{0}$ generate translations in $x_{n}^{0}$ which means $[x_{n}^{0},(p_{n})_{0}]=i$. Therefore we have
\begin{equation}\label{unitay}
	|\acute{t}_{n}\rangle=e^{-i(\acute{t}-t)(p_{n})_{0}}|t_{n}\rangle,
\end{equation}
where $|t_{n}\rangle$ and $|\acute{t}_{n}\rangle$ are eigenkets of operator $x_{n}^{0}$ with eigenvalues $t$ and $\acute{t}$, respectively.

The next step is obtaining Schrödinger equation for conditional state $|\psi_{s}(t)\rangle$ in curved space-time(Rindler background) by using PaW approach. In doing so we do as follows
\begin{equation}\label{tensor2}
	i\frac{d}{dt}|\psi_{s}(t)\rangle=i\frac{d}{dt}\left(\otimes_{n}\langle t_{n}|\otimes I_{S}\right)|\Psi\rangle\rangle=\left(\sum_{m}\langle t|(p_{m})_{0}\otimes I_{\bar{m}}^{C}\otimes I_{S}\right)|\Psi\rangle\rangle,
\end{equation}
where $I_{\bar{m}}^{C}$ indicates the identity operator on all of Hilbert space $\mathcal{H}_{n}^{T}$ for which $n\not=m$. The equation $(\ref{unitay})$ also is used to calculate the derivative with respect to $t$. Then by applying constraint $C_{n}^{+}|\Psi\rangle\rangle=0$ and equation $(\ref{cplus})$ one can write
\begin{equation}\label{tensor1}
	(p_{m})_{0}\otimes I_{\bar{m}}^{C}\otimes I_{S}|\Psi\rangle\rangle = -I^{C} \otimes \tilde{h}_{\rho m} \otimes I_{S_{\bar{m}}}|\Psi\rangle \rangle,
\end{equation}
for all $m$ where $\tilde{h}_{\rho m} \in \mathcal{H}_{m}^{cm}\otimes \mathcal{H}_{m}^{clock}$ and is obtained from operator equivalent of equation $(\ref{hro})$. The identity operator $I_{S_{\bar{m}}}$ acting on $\otimes_{n\not=m}\mathcal{H}_{m}^{cm}\otimes \mathcal{H}_{m}^{clock}$. By inserting equation $(\ref{tensor1})$ into $(\ref{tensor2})$ we have
\begin{equation}
	i\frac{d}{dt}|\psi_{s}(t)\rangle\rangle= \left(\sum_{m}\langle t| \otimes \tilde{h}_{\rho m} \otimes I_{S_{\bar{m}}}\right)|\Psi\rangle\rangle = \sum_{m}\tilde{h}_{\rho m} \otimes I_{S_{\bar{m}}} |\psi_{s}(t)\rangle,
\end{equation}
which leads to the Schrödinger equation as follows
\begin{equation}\label{sch}
	i\frac{d}{dt}|\psi_{s}(t)\rangle = H_{S}|\psi_{s}(t)\rangle,
\end{equation}
where the Hamiltonian is $H_{S}= \sum_{m} \tilde{h}_{\rho m} \otimes I_{S_{\bar{m}}}$. By inserting equation $(\ref{hro})$ into equation $(\ref{sch})$ we reach to the Schrödinger equation for $N$ relativistic  particles in Rindler curved space-time
\begin{equation}
i \frac{d}{dt}|\psi_{s}(t)\rangle=\sum_{n}\frac{1}{\rho}(m_{n}c^{2} + H_{n}^{cm}+ H_{n}^{clock}-\frac{1}{m_{n}c^{2}}[H_{n}^{cm}\otimes H_{n}^{clock} + \frac{1}{2}(H_{n}^{cm})^{2}])\otimes I_{S_{\bar{n}}}|\psi_{s}(t)\rangle,
\end{equation}
then by redefining Hamiltonians with new Hamiltonians the above equation leads to
\begin{equation}\label{schrho}
i \frac{d}{dt}|\psi_{s}(t)\rangle=\sum_{n}(\tilde{H}_{n}^{0} + \tilde{H}_{n}^{cm}+ \tilde{H}_{n}^{clock}-\frac{\rho}{m_{n}c^{2}}[\tilde{H}_{n}^{cm}\otimes \tilde{H}_{n}^{clock} + \frac{1}{2}(\tilde{H}_{n}^{cm})^{2}])\otimes I_{S_{\bar{n}}}|\psi_{s}(t)\rangle,
\end{equation}
where we have used the
\begin{equation}
\tilde{H}_{n}^{clock}=\frac{H_{n}^{clock}}{\rho},
\end{equation}
\begin{equation}
\tilde{H}_{n}^{cm}= H_{n}^{cm}/\rho,
\end{equation}
\begin{equation}\label{basis}
\tilde{H}_{n}^{0}=m_{n}c^{2}/\rho.
\end{equation}
We used the term new Hamiltonian for the tilde Hamiltonian $\tilde{H}$ because when one ignores the $\rho$ can reach Hamiltonians in flat background. Note that one cannot ignore the $\tilde{H}_{n}^{0}$ like as flat space-time because is not constant. Therefore in considering the time evolution of the system which is done in the next sections, one must insert the influence of this term too. The interaction Hamiltonian is given by
\begin{equation}\label{interactionrho}
\tilde{H}_{n}^{int}=-\frac{\rho}{m_{n}c^{2}}[\tilde{H}_{n}^{cm}\otimes \tilde{H}_{n}^{clock} + \frac{1}{2}(\tilde{H}_{n}^{cm})^{2}].
\end{equation}
The equation $(\ref{schrho})$ is the Schrödinger equation on conditional state $|\psi_{s}(t)\rangle$ in Rindler background for near horizon region of black hole deriving from timeless PaW mechanism.

\subsection{PaW mechanism by tortoise coordinate}
In this part, the PaW approach is considered in a background which is obtained by making the Rindler metric $(\ref{metric})$ conformal. In doing so we use the so-called tortoise coordinate which is so useful when one interested in field theory in the near-horizon region of black holes. In doing so we put $\ln(\rho)=z$ leading to the following line element,
\begin{equation}\label{tor}
ds^{2}=	e^{2z}(-d\tau ^{2}+dz^{2}).
\end{equation}
Note that the location of the black hole's horizon is $\rho =0$ which in tortoise coordinate becomes $z \rightarrow -\infty$. Therefore in this coordinate one can stretch the near horizon region, especially, when is close to the horizon's location.

We repeat the previous section's calculations in this part and then derive the Schrödinger equation in the background of metric $(\ref{tor})$. In doing so we apply metric $(\ref{tor})$ into equation $(\ref{ch})$ leading to
\begin{equation}
	C_{H_{n}}=-e^{2z}(p_{n}^{0})^{2}c^{2}+(p_{n}^{1})^{2}c^{2}e^{2z}+(m_{n}+\frac{H_{n}^{clock}}{c^{2}})^{2}c^{4} \approx 0.
\end{equation}
Manipulating the above equation yields
\begin{equation}
\sqrt{(p_{n}^{0})^{2}c^{2}}=\sqrt{(p_{n}^{1})^{2}c^{2}+e^{-2z}(m_{n}+H_{n}^{clock}/c^{2})^{2}c^{4}}=h_{zn}.
\end{equation}
In this case $C_{n}^{+}$ can be defined as
\begin{equation}\label{cpluszed}
C_{n}^{+}= (p_{n})_{0}+h_{z n}.
\end{equation}
Now we expand $h_{zn}$ by ignoring the quadratic interacting terms as follows
\begin{equation}
h_{zn}=m_{n}c^{2}e^{-z}[1+\frac{(p_{n}^{1})^{2}}{2m_{n}^{2}c^{2}e^{-2z}}+\frac{H_{n}^{clock}}{m_{n}c^{2}}-
\frac{(p_{n}^{1})^{2}H_{n}^{clock}}{2m_{n}^{3}c^{4}e^{-2z}}-\frac{(H_{n}^{cm})^{2}}{2e^{-4z}(m_{n}c^{2})^{2}}].
\end{equation}
By manipulating the above equation we have
\begin{equation}\label{hzn}
h_{zn}=m_{n}c^{2}e^{-z}+\frac{H_{n}^{cm}}{e^{-z}}+H_{n}^{clock}e^{-z}-\frac{H_{n}^{cm}\otimes H_{n}^{clock}}{m_{n}c^{2}e^{-z}}-\frac{(H_{n}^{cm})^{2}}{2m_{n}c^{2}e^{-3z}}+ ...
\end{equation}
If we follow the logic of the previous subsection for deriving the Schrödinger equation in the Rindler background we reach the conclusion that the equation $(\ref{tensor2})$ is the same in conformal metric $(\ref{tor})$. Then by applying constraint $C_{n}^{+}|\Psi\rangle\rangle=0$ and equation $(\ref{cpluszed})$ one can write
\begin{equation}\label{tensor3}
(p_{m})_{0}\otimes I_{\bar{m}}^{C}\otimes I_{S}|\Psi\rangle\rangle = -I^{C} \otimes h_{z m} \otimes I_{S_{\bar{m}}}|\Psi\rangle \rangle,
\end{equation}
where it is correct for all $m$ and $h_{z m} \in \mathcal{H}_{m}^{cm}\otimes \mathcal{H}_{m}^{clock}$ which is obtained from operator equivalent of equation $(\ref{hzn})$. The identity operator $I_{S_{\bar{m}}}$ acting on $\otimes_{n\not=m}\mathcal{H}_{m}^{cm}\otimes \mathcal{H}_{m}^{clock}$. By using equation $(\ref{tensor3})$ in $(\ref{tensor2})$ it is given that
\begin{equation}
i\frac{d}{dt}|\psi_{s}(t)\rangle\rangle= \left(\sum_{m}\langle t| \otimes h_{z m} \otimes I_{S_{\bar{m}}}\right)|\Psi\rangle\rangle = \sum_{m}h_{z m} \otimes I_{S_{\bar{m}}} |\psi_{s}(t)\rangle,
\end{equation}
which, in fact, is the Schrödinger equation as follows
\begin{equation}\label{schzn}
i\frac{d}{dt}|\psi_{s}(t)\rangle = H_{S}|\psi_{s}(t)\rangle,    ~~~~ with ~~~~  H_{S}= \sum_{m} h_{z m} \otimes I_{S_{\bar{m}}}.
\end{equation}
Now by inserting equation $(\ref{hzn})$ into $(\ref{schzn})$ the Schrödinger equation of $N$ relativistic particles in background of metric $(\ref{tor})$ is given by
\begin{eqnarray}i\frac{d}{dt}|\psi_{s}(t)\rangle=
\sum_{n}(m_{n}c^{2}e^{-z}+\frac{H_{n}^{cm}}{e^{-z}}+H_{n}^{clock}e^{-z}
-\frac{H_{n}^{cm}\otimes H_{n}^{clock}}{m_{n}c^{2}e^{-z}}\\\nonumber
-\frac{(H_{n}^{cm})^{2}}{2m_{n}c^{2}e^{-3z}})\otimes I_{S_{\bar{n}}}|\psi_{s}(t)\rangle,
\end{eqnarray}
by manipulating above equation the following equation is obtained
\begin{equation}\label{schhnz}
i \frac{d}{dt}|\psi_{s}(t)\rangle=
\sum_{n}(H_{zn}^{0}
+ H_{zn}^{cm}+ H_{zn}^{clock}
-\frac{e^{z}}{m_{n}c^{2}}[H_{zn}^{cm}\otimes H_{zn}^{clock} + \frac{1}{2}(H_{zn}^{cm})^{2}])\otimes I_{S_{\bar{n}}}|\psi_{s}(t)\rangle,
\end{equation}
where we have used the following variations
\begin{equation}
	H_{nz}^{0}=m_{n}c^{2}e^{-z}=H_{n}^{0}e^{-z},
\end{equation}
\begin{equation}
	H_{nz}^{cm}=\frac{H_{n}^{cm}}{e^{-z}},
\end{equation}
\begin{equation}
	H_{nz}^{clock}=H_{n}^{clock}e^{-z}.
\end{equation}
Note that $(nz)$ is only an index and there is no difference between $(nz)$ and $(zn)$. The interaction term is
\begin{equation}\label{interationzn}
	H_{nz}^{int}=-\frac{e^{z}}{m_{n}c^{2}}\left(H_{nz}^{cm}\otimes H_{nz}^{clock}+\frac{1}{2}(H_{nz}^{cm})^{2}\right).
\end{equation}
The equation $(\ref{schhnz})$ is the Schrödinger equation on conditional state $|\psi_{s}(t)\rangle$ in conformal background by metric $(\ref{tor})$ for near horizon region of black hole deriving from timeless PaW mechanism.

\section{Quantum time dilation in Rindler background in near horizon region }
In this section, we consider two particles $A$ and $B$ with an internal degree of freedom interpreted as a clock. We suppose that these two particles $A$ and $B$ are moving in curved space-time which supported by $(\ref{metric})$ and quantum properties as $\{\mathcal{H}_{A}^{clock}, |\psi_{A}^{clock}\rangle, H_{A}^{clock}, T_{A}\}$ and $\{\mathcal{H}_{A}^{clock}, |\psi_{B}^{clock}\rangle, H_{B}^{clock}, T_{B}\}$, respectively. The clock moving in Rindler space describing by a physical state $|\Psi\rangle\rangle$. The quantum features of these clocks are governed by the following equation
\begin{equation}\label{wheelertotal}
(\hslash P_{T}\otimes I_{S}+I_{S}\otimes H_{S})|\Psi\rangle\rangle=0,
\end{equation}
which is obtained by using equations $(\ref{hcrossh})$ and $(\ref{wheeler})$. Each particle is equipped by equation $(\ref{wheelertotal})$ which in a brief form can be written as $(H_{C}+H_{S})|\Psi\rangle\rangle=0$ with $H_{C}=\hslash P_{T}$ where $H_{C}\in \mathcal{H}_{T}$ and $H_{S} \in \mathcal{H}_{S}$ which have defined in the previous sections as local Hamiltonians of "Clock" and "system", receptively. A time observable $T$ in Hilbert space $\mathcal{H}_{T}$ can be defined as POVM.

Now we want to calculate the time dilation effects between clocks $A$ and $B$. In doing so, one can consider the probability that clock $A$ reads the proper time $\tau_{A}$ conditioned on clock $B$ reading the proper time $\tau_{B}$. This conditional probability is as follows \cite{main}.
\begin{eqnarray}\label{probrho}
Prob[T_{A}=\tau_{A}|T_{B}=\tau_{B}]
=\frac{Prob[T_{A}=\tau_{A} \& T_{B}=\tau_{B}]}{Prob[T_{B}=\tau_{B}]}\\\nonumber
=\frac{\langle\langle \Psi|E_{A}(\tau_{A})E_{B}(\tau_{B})|\Psi\rangle\rangle}{\langle\langle \Psi|E_{B}(\tau_{B})|\Psi\rangle\rangle},
\end{eqnarray}
where $E_{A}(\tau_{A})$ and $E_{B}(\tau_{B})$ are the effect operators for clock $A$ and $B$ which define $T_{clock}$ and are proportional to "projection" operators
\begin{equation}
	E_{clock}(\tau)= \zeta |\tau\rangle\langle \tau|,
\end{equation}
for $\zeta \in R$ and $\int_{G} d\tau E_{clock}(\tau)=I_{clock} $ where $G$ is the group generated by $H_{clock}$.
To derive the probability distribution by noting that the "Clock" state $E_{C}(t)=|t\rangle\langle t|$ form a basis dense in $\mathcal{H}_{T}$ one can expand the physical state as follows
\begin{equation}
	|\Psi\rangle\rangle=\int dt |t \rangle |\psi_{s}(t)\rangle=\int dt |t\rangle\otimes_{n \in \{A,B\}}U_{n}(t)|\psi_{n}^{cm}\rangle|\psi_{n}^{clock}\rangle,
\end{equation}
where $U_{n}(t)=exp(-iH_{n}t)$ and $H_{n}$ is given by equation $(\ref{hro})$. Note that at the starting point $t=0$ we have $|\psi_{s}(0)\rangle=|\psi_{A}\rangle|\psi_{B}\rangle$ which is an unentangled state. Also we suppose that the center-of-mass and internal clock degrees of freedom are not entangled $|\psi_{n}\rangle=|\psi_{n}^{cm}\rangle|\psi_{n}^{clock}\rangle$. To calculate conditional probability $(\ref{probrho})$ we do as follows.

The leading order relativistic contribution in Rindler background in near horizon region is given by equation $(\ref{interactionrho})$ and the Hamiltonian in equation $(\ref{hro})$ can be expressed as
\begin{equation}
	\tilde{h}_{\rho n}=\tilde{H}_{n}^{0} + \tilde{H}_{n}^{clock}+\tilde{H}_{n}^{cm}+\tilde{H}_{n}^{int}.
\end{equation}
At first, we find the free evolution of the center-of-mass, internal clock degrees of freedom, and the term coming from rest mass as follows
\begin{equation}
\rho_{n}^{cm}(t)=e^{-i\tilde{H}_{n}^{cm}t}\rho_{n}^{cm}e^{i\tilde{H}_{n}^{cm}t},
\end{equation}
\begin{equation}
\rho_{n}^{clock}(t)=e^{-i\tilde{H}_{n}^{clock}t}\rho_{n}e^{i\tilde{H}_{n}^{clock}t},
\end{equation}
\begin{equation}\label{constant}
\rho_{n}^{0}(t)=e^{-i\tilde{H}_{n}^{0}t}\rho_{n}e^{i\tilde{H}_{n}^{0}t},
\end{equation}
for $n\in \{A,B\}$ where labels the clocks $A$ and $B$. Not that we cannot ignore the term $(\ref{constant})$ coming from rest mass since it is not constantly adding to total Hamiltonian. Therefore the reduced state of the clock to leading Rindler relativistic space order is given by
\begin{equation}
\rho_{n}(t)=tr_{cm,0}(e^{-i\tilde{H}_{n}^{int}t}\rho_{n}^{0}(t)\otimes\rho_{n}^{cm}(t)\otimes\rho_{n}^{clock}(t)e^{i\tilde{H}_{n}^{int}t}),
\end{equation}
which leads to
\begin{equation}\label{trcm}
\rho_{n}(t)=\rho_{n}^{clock}(t)+it\frac{\langle\rho\tilde{H}_{n}^{cm}\rangle}{mc^{2}}[\tilde{H}_{n}^{clock},\rho_{n}^{clock}(t)].
\end{equation}
To calculate the conditional probability distribution $(\ref{probrho})$ it is needed to evaluate following relation
\begin{eqnarray}\label{perturb}
tr(E_{n}(\tau_{n})\rho_{n}(t))=\\\nonumber
&tr(E_{n}(\tau_{n})\rho_{n}^{clock}(t))+
it\frac{\langle\rho\tilde{H}_{n}^{cm}\rangle}{mc^{2}}tr(E_{n}(\tau_{n})[\tilde{H}_{n}^{clock},\rho_{n}^{clock}(t)]).
\end{eqnarray}
The equation $(\ref{perturb})$ leads to a solution perturbatively, for conditional probability $(\ref{probrho})$. We suppose the fiducial state $|\psi_{n}^{clock}\rangle \in \mathcal{H}_{n}^{clock}$ is Gaussian with a spread $\sigma$, then the first term in equation $(\ref{perturb})$ is expressed as
\begin{eqnarray}\label{fisrtperturb}
tr(E_{n}(\tau_{n})\rho_{n}^{clock}(t))= \left|\langle\tau_{n}|e^{-i\tilde{H}_{n}^{clock}t}|\psi_{n}^{clock}\rangle\right|^{2}\\\nonumber
= \left|\int_{R} d\acute{\tau}\langle\tau_{n}|\acute{\tau}_{n}\rangle\frac{e^{-\frac{(\acute{\tau}-t)^{2}}{2\sigma^{2}}}}{\pi^{\frac{1}{4}}\sqrt{\sigma}} \right|^{2} \\\nonumber
=\frac{e^{-\frac{(\tau-t)^{2}}{\sigma^{2}}}}{\pi^{\frac{1}{2}}\sigma},
\end{eqnarray}
where we applied the orthogonality of the clock states as $\langle\tau_{n}|\acute{\tau}_{n}\rangle=\delta(\tau-\acute{\tau})$ which for an ideal clock is a suitable choice. To calculate the second term in equation $(\ref{perturb})$, by defining $|\psi_{n}^{clock}(t)\rangle:=e^{-i\tilde{H}_{n}^{clock}t}|\psi_{n}^{clock}\rangle$ the following trace is given by
\begin{eqnarray}\label{simple}
tr\left[E(\tau_{n})[\tilde{H}_{n}^{clock},\rho_{n}^{clock}] \right]
=\langle\tau_{n}|[\tilde{H}_{n}^{clock},\rho_{n}^{clock}]|\tau_{n}\rangle\\\nonumber
=\langle\tau_{n}|\tilde{H}_{n}^{clock}|\psi_{n}^{clock}(t)\rangle\langle\psi_{n}^{clock}(t)|\tau_{n}\rangle\\\nonumber
-\langle\tau_{n}|\psi_{n}^{clock}(t)\rangle\langle\psi_{n}^{clock}(t)|\tilde{H}_{n}^{clock}|\tau_{n}\rangle \\\nonumber
=\left(\langle\tau_{n}|\tilde{H}_{n}^{clock}|\psi_{n}^{clock}(t)\rangle
-\langle\psi_{n}^{clock}(t)|\tilde{H}_{n}^{clock}|\tau_{n}\rangle \right)\frac{e^{-\frac{(\tau-t)^{2}}{2\sigma^{2}}}}{\pi^{\frac{1}{4}}\sqrt{\sigma}}.
\end{eqnarray}
By noting that the clock states satisfy
\begin{equation}
|(\tau + \acute{\tau})_{n}\rangle=e^{-i\tilde{H}_{n}^{clock}\acute{\tau}}|\tau_{n}\rangle.
\end{equation}
Implying $H_{n}^{clock}=-i\partial/\partial\tau$ which is the displacement operator in $|\tau_{n}\rangle$ representation \cite{displace}. Then we have
\begin{equation}
\langle\tau_{n}|\tilde{H}_{n}^{clock}|\psi_{n}^{clock}(t)\rangle=-i\frac{\partial}{\partial \tau}\frac{e^{-\frac{(\tau-t)^{2}}{2\sigma^{2}}}}{\pi^{\frac{1}{4}}\sqrt{\sigma}}=i\frac{e^{-\frac{(\tau-t)^{2}}{2\sigma^{2}}}}
{\pi^{\frac{1}{4}}\sqrt{\sigma}}\frac{\tau - t}{\sigma^{2}}.
\end{equation}
Therefore the equation $(\ref{simple})$ is simplified as
\begin{equation}\label{simple1}
tr\left[E(\tau_{n})[\tilde{H}_{n}^{clock},\rho_{n}^{clock}] \right]=2i\frac{e^{-\frac{(\tau-t)^{2}}{\sigma^{2}}}}{\pi^{\frac{1}{2}}\sigma}\frac{\tau - t}{\sigma^{2}}.
\end{equation}
Finally by using equations $(\ref{simple1})$,$(\ref{simple})$ and $(\ref{fisrtperturb})$ with equation $(\ref{perturb})$ together we have
\begin{eqnarray}\label{finalresult}
tr(E_{n}(\tau_{n})\rho_{n}(t))
=tr(E_{n}(\tau_{n})\rho_{n}^{clock}(t))\\\nonumber + it\frac{\langle\rho\tilde{H}_{n}^{cm}\rangle}{mc^{n}}tr(E_{n}(\tau_{n})[\tilde{H}_{n}^{clock},\rho_{n}^{clock}(t)])\\\nonumber
=\frac{e^{-\frac{(\tau-t)^{2}}{\sigma^{2}}}}{\pi^{\frac{1}{2}}\sigma}+ it\frac{\langle\rho \tilde{H}_{n}^{cm}\rangle}{mc^{2}}\left(2i\frac{e^{-\frac{(\tau-t)^{2}}{\sigma^{2}}}}{\pi^{\frac{1}{2}}\sigma}\frac{\tau - t}{\sigma^{2}}\right)\\\nonumber
=\frac{e^{-\frac{(\tau-t)^{2}}{\sigma^{2}}}}{\pi^{\frac{1}{2}}\sigma^{2}}\left[1-2\frac{\langle\rho \tilde{H}_{n}^{cm}\rangle}{mc^{2}}\frac{t(\tau-t)}{\sigma^{2}}\right].
\end{eqnarray}
Then by using equation $(\ref{finalresult})$ in equation $(\ref{probrho})$ the conditional probability is
\begin{eqnarray}\label{probrhofinal}
Prob[T_{A}=\tau_{A}|T_{B}=\tau_{B}]=\frac{\int dt ~tr[E_{A}(\tau_{A})\rho_{A}(t)]tr[E_{A}(\tau_{B})\rho_{B}(t)]}{\int dt ~tr[E_{A}(\tau_{B})\rho_{B}(t)]}\\\nonumber
=\frac{e^{-\frac{(\tau_{A}-\tau_{B})^{2}}{2\sigma^{2}}}}{(2\pi)^{\frac{1}{2}}\sigma}\frac{1+\frac{\langle\rho \tilde{H}_{A}^{cm}\rangle+\langle\rho \tilde{H}_{B}^{cm}\rangle}{2mc^{2}}-\frac{\langle\rho \tilde{H}_{A}^{cm}\rangle-\langle\rho \tilde{H}_{B}^{cm}\rangle}{2mc^{2}}\frac{\tau_{A}^{2}-\tau_{B}^{2}}{\sigma^{2}}}{1+\frac{\langle\rho \tilde{H}_{B}^{cm}\rangle}{mc^{2}}}\\\nonumber
=\frac{e^{-\frac{(\tau_{A}-\tau_{B})^{2}}{2\sigma^{2}}}}{(2\pi)^{\frac{1}{2}}\sigma}\left(1+ \frac{\langle\rho \tilde{H}_{A}^{cm}\rangle-\langle\rho \tilde{H}_{B}^{cm}\rangle}{mc^{2}}\frac{\sigma^{2}-\tau_{A}^{2}+\tau_{B}^{2}}{2\sigma^{2}}\right).	
\end{eqnarray}

To derive conditional probability $(\ref{probrhofinal})$ in near horizon region background (Rindler) explicitly, we calculate $\langle\rho \tilde{H}_{n}^{cm}\rangle$ as follows
\begin{equation}
\langle\rho \tilde{H}_{n}^{cm}\rangle=\langle\psi_{n}^{0}|\langle\psi_{n}^{cm}|\rho \tilde{H}_{n}^{cm}|\psi_{n}^{cm}\rangle|\psi_{n}^{0}\rangle=\rho \langle\psi_{n}^{cm}|\tilde{H}_{n}^{cm}|\psi_{n}^{cm}\rangle.
\end{equation}
Also note that
\begin{equation}
\tilde{H}_{n}^{cm}=\frac{{H_{n}^{cm}}}{\rho}=\frac{p_{n}^{2}}{2m\rho}=\frac{\mathcal{P}_{n\rho}^{2}}{2m}
\end{equation}
where $(\mathcal{P}_{n\rho})^{2}=\frac{g_{ij}p_{n}^{i}p_{n}^{j}}{\rho}$ with $(i,j = 1,2,3)$.

As we described the conditional probability, the average proper time read by clock $A$ conditional on clock $B$ indicating the time $\tau_{B}$ in the near horizon region background (Rindler space) is given by \footnote{Note that according to "Desiderata of physical clocks theorem" if $T_{clock}$ is a covariant time observable relative to group generated by $H_{clock}$ and a fiducial state $\rho$ such that $\langle T_{clock}\rangle_{\rho}=0$ and $\rho(\tau):=U_{clock}(\tau)\rho U_{clock}^{\dagger}(\tau)$ then $T_{clock}$ is an unbiased estimator of the parameter $\tau$ such that $\langle T_{clock}\rangle_{\rho(\tau)}=\tau$. For more details refer to \cite{main}. However it is important to note that the relation $\langle T_{clock}\rangle_{\rho(\tau)}=\tau$ is correct independent of background space-time. In other words, in calculating the average of $T_{clock}$ on the state $\rho(\tau)$ the result does not depend on background space-time. Therefore in the Rindler space background and also the conformal case that we will consider in the next section, we have $\langle T_{clock}\rangle_{\rho(\tau)}=\tau$. }
\begin{equation}\label{dilation}
\langle T_{A}\rangle=\left(1-\frac{\langle\rho \tilde{H}_{A}^{cm}\rangle-\langle\rho \tilde{H}_{B}^{cm}\rangle}{mc^{2}}\right)\tau_{B}.
\end{equation}
Therefore according to the above considerations and the footnote below this page, the relation $(\ref{dilation})$ is the observed average time dilation between clocks $A$ and $B$. To consider the explicit form of equation $(\ref{dilation})$  we suppose the center-of-mass of both clocks is equipped with a Gaussian state which is localized in around an average momentum $\bar{\mathcal{P}}_{n\rho}$ with spread $\Delta _{n}>0$
\begin{equation}
|\psi_{n}^{cm}\rangle=\frac{1}{\pi^{1/4}\sqrt{\Delta_{n}}}\int d\mathcal{P}exp(\frac{-(\mathcal{P}-\bar{\mathcal{P}}_{n\rho})^{2}}{2\Delta_{n}^{2}})|\mathcal{P}_{n\rho}\rangle.
\end{equation}
By using this state to consider average amount we have
\begin{equation}\label{average}
\langle\rho \tilde{H}_{n}^{cm}\rangle=\rho \frac{\bar{\mathcal{P}}_{n\rho}^{2}}{2m}+\rho \frac{\Delta_{n}^{2}}{4m}.
\end{equation}
Finally by applying $(\ref{average})$ into equation $(\ref{dilation})$ the explicit form of average time dilation between two clocks $A$ and $B$ in near horizon region of black hole describing by Rindler metric $(\ref{metric})$ is
\begin{equation}\label{dilationrho}
\langle T_{A}\rangle=\left[1-\rho \frac{\bar{\mathcal{P}}_{A\rho}^{2}-\bar{\mathcal{P}}_{B\rho}^{2}+\frac{1}{2}(\Delta_{A}^{2}-\Delta_{B}^{2})}{2m^{2}c^{2}}\right]\tau_{B}.
\end{equation}
In the classical limit $\langle\rho \tilde{H}_{n}^{cm}\rangle \rightarrow \frac{\bar{p}_{n}^{2}}{2m}$ where the $\Delta_{n}$ is ignored and $\bar{p}_{n}^{2}$ corresponding to the average velocity of momentum wave packets of the clock. Then to leading relativistic order the proper time $\tau_{A}$ recognized by $A$ when $B$ reads the proper time $\tau_{B}$ is given by
\begin{equation}\label{classicaldilation}
	\tau_{A}=\frac{\gamma_{B}}{\gamma_{A}}\tau_{B}=[1-\frac{\bar{p}_{A}^{2}-\bar{p}_{B}^{2}}{2m^{2}c^{2}}]\tau_{B},
\end{equation}
where $\gamma_{n}:=\sqrt{1+\frac{\bar{p}_{n}^{2}}{m^{2}c^{2}}}$. Therefore by ignoring the quantum feature of the clocks, we reach to the classical equation for time dilation, on the other hand, the curved space-time in the appropriate limit gives a flat space limit for time dilation.

It is important to note that the time dilation equation $(\ref{dilationrho})$ when one is close to the horizon of black hole $(\rho \rightarrow 0)$ leads to $\tau_{A}=\tau_{B}$ which asserts that for there is no time dilation in the horizon of the black hole in consistent with horizon feature of the black hole.

\section{Quantum time dilation in near horizon region by tortoise coordinate }

We have all of the interpretations that we had in the previous section for the near horizon region of the black hole by tortoise coordinate too. To find Time dilation between clocks for this case, we do as follows. According to the previous section, we obtain free evolution of the center-of-mass, internal clock degrees of freedom, and the term coming from rest mass in the near horizon region supported by conformal metric $(\ref{tor})$ as follows
\begin{equation}
\rho_{nz}^{cm}(t)=e^{-iH_{nz}^{cm}t}\rho_{nz}^{cm}e^{iH_{nz}^{cm}t},
\end{equation}
\begin{equation}
\rho_{nz}^{clock}(t)=e^{-iH_{nz}^{clock}t}\rho_{nz}e^{iH_{nz}^{clock}t},
\end{equation}
\begin{equation}\label{constantzn}
\rho_{nz}^{0}(t)=e^{-iH_{nz}^{0}t}\rho_{nz}e^{iH_{nz}^{0}t},
\end{equation}
where $n \in {A,B}$ are the indicator of clocks $A$ and $B$. In this case, we have to consider the state  $(\ref{constantzn})$ and Hamiltonian $H_{nz}^{0}$ because it is not like a constant adding to the total energy of the system. Following the same strategy of the previous section the leading order relativistic contribution in near horizon region supported by metric $(\ref{tor})$ is given by equation $(\ref{interationzn})$ and the Hamiltonian in equation $(\ref{hzn})$ can be expressed as
\begin{equation}
	h_{nz}=H_{nz}^{0}+H_{nz}^{cm}+H_{nz}^{clock}+H_{nz}^{int}.
\end{equation}
Then, we find the reduced state as
\begin{equation}
\rho_{nz}(t)=tr_{cm,0}(e^{-iH_{nz}^{int}t}\rho_{nz}^{0}(t)\otimes\rho_{nz}^{cm}(t)
\otimes\rho_{nz}^{clock}(t)e^{iH_{nz}^{int}t}),
\end{equation}
by manipulating the above equation we get
\begin{equation}
\rho_{nz}(t)=\rho_{nz}^{clock}(t)+it\frac{\langle e^{z} H_{nz}^{cm}\rangle}{mc^{2}}[H_{nz}^{clock},\rho_{nz}^{clock}(t)].
\end{equation}
To find conditional probability distribution we do as follows. Like the previous section we have
\begin{eqnarray}\label{ansatz}
tr(E_{nz}(\tau_{n})\rho_{nz}(t))\\\nonumber
=&tr(E_{nz}(\tau_{n})\rho_{nz}^{clock}(t)) + it\frac{\langle e^{z} H_{nz}^{cm}\rangle}{mc^{2}}tr(E_{n}(\tau_{n})[H_{nz}^{clock},\rho_{nz}^{clock}(t)]).
\end{eqnarray}
The equation $(\ref{ansatz})$ leads to a solution perturbatively, for conditional probability
$(\ref{probrho})$ for the case of relativistic particles supported by metric $(\ref{tor})$. The fiducial state $|\psi_{n}^{clock}\rangle \in \mathcal{H}_{n}^{clock}$ is supposed to be Gaussian with a spread $\sigma$. Then the first term in equation $(\ref{ansatz})$ results in
\begin{eqnarray}
tr(E_{nz}(\tau_{n})\rho_{nz}^{clock}(t))&=&\left|\langle\tau_{n}|e^{-i H_{nz}^{clock}t}|\psi_{n}^{clock}\rangle\right|^{2}\\\nonumber
&=&\left|\int_{R} d\acute{\tau}\langle\tau_{n}|\acute{\tau}_{n}\rangle\frac{e^{-\frac{(\acute{\tau}-t)^{2}}{2\sigma^{2}}}}{\pi^{\frac{1}{4}}\sqrt{\sigma}} \right|^{2} \\\nonumber
&=&\frac{e^{-\frac{(\tau-t)^{2}}{\sigma^{2}}}}{\pi^{\frac{1}{2}}\sigma},
\end{eqnarray}
where we used orthogonality of ideal clock states as $\langle\tau_{n}|\acute{\tau}_{n}\rangle=\delta(\tau-\acute{\tau})$.
In order to calculate the second term of equation $(\ref{ansatz})$, by defining $|\psi_{n}^{clock}(t)\rangle:=e^{-iH_{nz}^{clock}t}|\psi_{n}^{clock}\rangle$ the following trace is
\begin{eqnarray}\label{longtrace}
tr\left[E(\tau_{nz})[H_{nz}^{clock},\rho_{nz}^{clock}] \right]
=<\tau_{n}|[H_{nz}^{clock},\rho_{nz}^{clock}]|\tau_{n}>\\\nonumber
=<\tau_{n}|H_{nz}^{clock}|\psi_{n}^{clock}(t)><\psi_{n}^{clock}(t)|\tau_{n}>\\\nonumber
-<\tau_{n}|\psi_{n}^{clock}(t)><\psi_{n}^{clock}(t)|H_{nz}^{clock}|\tau_{n}>\\\nonumber
=\left(<\tau_{n}|H_{nz}^{clock}|\psi_{n}^{clock}(t)>
-<\psi_{n}^{clock}(t)|H_{nz}^{clock}|\tau_{n}> \right)\frac{e^{-\frac{(\tau-t)^{2}}{2\sigma^{2}}}}{\pi^{\frac{1}{4}}\sqrt{\sigma}}.
\end{eqnarray}
By applying displacement operator $H_{nz}^{clock}=-i\frac{\partial}{\partial \tau}$ in $|\tau_{n}\rangle$ representation and using
\begin{equation}
|(\tau + \acute{\tau})_{n}\rangle=e^{-iH_{nz}^{clock}\acute{\tau}}|\tau_{n}\rangle.
\end{equation}
Then we have
\begin{equation}
\langle\tau_{n}|H_{nz}^{clock}|\psi_{n}^{clock}(t)\rangle=-i\frac{\partial}{\partial \tau}\frac{e^{-\frac{(\tau-t)^{2}}{2\sigma^{2}}}}{\pi^{\frac{1}{4}}\sqrt{\sigma}}=i\frac{e^{-\frac{(\tau-t)^{2}}{2\sigma^{2}}}}{\pi^{\frac{1}{4}}\sqrt{\sigma}}\frac{\tau - t}{\sigma^{2}}.
\end{equation}
By considering the above results equation $(\ref{longtrace})$ is rewritten as follows
\begin{equation}\label{traceresult}
tr\left[E(\tau_{n})[H_{nz}^{clock},\rho_{nz}^{clock}] \right]=2i\frac{e^{-\frac{(\tau-t)^{2}}{\sigma^{2}}}}{\pi^{\frac{1}{2}}\sigma}\frac{\tau - t}{\sigma^{2}}.
\end{equation}
Then the final result for equation $(\ref{ansatz})$ becomes
\begin{eqnarray}\label{finalzn}
tr(E_{nz}(\tau_{n})\rho_{nz}(t))=tr(E_{nz}(\tau_{n})\rho_{nz}^{clock}(t))\\\nonumber
 + it\frac{\langle e^{z} H_{nz}^{cm}\rangle}{mc^{2}}tr(E_{nz}(\tau_{n})[H_{nz}^{clock},\rho_{nz}^{clock}(t)])\\\nonumber
=\frac{e^{-\frac{(\tau-t)^{2}}{\sigma^{2}}}}{\pi^{\frac{1}{2}}\sigma}+ it\frac{\langle e^{z} H_{nz}^{cm}\rangle}{mc^{2}}\left(2i\frac{e^{-\frac{(\tau-t)^{2}}{\sigma^{2}}}}{\pi^{\frac{1}{2}}\sigma}\frac{\tau - t}{\sigma^{2}}\right)\\\nonumber
=\frac{e^{-\frac{(\tau-t)^{2}}{\sigma^{2}}}}{\pi^{\frac{1}{2}}\sigma^{2}}\left[1-2\frac{\langle e^{z} H_{nz}^{cm}\rangle}{mc^{2}}\frac{t(\tau-t)}{\sigma^{2}}\right].
\end{eqnarray}
Thus by applying equation $(\ref{finalzn})$ in $(\ref{probrho})$ conditional probability for the two clocks $A$ and $B$ in near horizon region supported by metric $(\ref{tor})$ is written as
\begin{eqnarray}\label{probzn}
Prob[T_{A}=\tau_{A}|T_{B}=\tau_{B}]=\frac{\int dt ~tr[E_{Az}(\tau_{A})\rho_{Az}(t)]tr[E_{Az}(\tau_{B})\rho_{Bz}(t)]}{\int dt ~tr[E_{Az}(\tau_{Bz})\rho_{Bz}(t)]}\\\nonumber
=\frac{e^{-\frac{(\tau_{A}-\tau_{B})^{2}}{2\sigma^{2}}}}{(2\pi)^{\frac{1}{2}}\sigma}\frac{1+\frac{\langle e^{z} H_{Az}^{cm}\rangle+\langle e^{z} H_{Bz}^{cm}\rangle}{2mc^{2}}-\frac{\langle e^{z} H_{Az}^{cm}\rangle-\langle e^{z} H_{Bz}^{cm}\rangle}{2mc^{2}}\frac{\tau_{A}^{2}-\tau_{B}^{2}}{\sigma^{2}}}{1+\frac{\langle e^{z} H_{Bz}^{cm}\rangle}{mc^{2}}}\\\nonumber
=\frac{e^{-\frac{(\tau_{A}-\tau_{B})^{2}}{2\sigma^{2}}}}{(2\pi)^{\frac{1}{2}}\sigma}\left(1+ \frac{\langle e^{z} H_{Az}^{cm}\rangle-\langle e^{z} H_{Bz}^{cm}\rangle}{mc^{2}}\frac{\sigma^{2}-\tau_{A}^{2}+\tau_{B}^{2}}{2\sigma^{2}}\right).	
\end{eqnarray}

To write conditional probability $(\ref{probzn})$ explicitly we derive $\langle e^{z}H_{nz}^{cm}\rangle$

\begin{equation}
	\langle e^{z}H_{nz}^{cm}\rangle=\langle\psi_{n}^{0}|\langle\psi_{n}^{cm}|e^{z}H_{nz}^{cm}|\psi_{n}^{cm}\rangle|\psi_{n}^{0}
\rangle=e^{z}\langle\psi_{n}^{cm}|H_{nz}^{cm}|\psi_{n}^{cm}\rangle.
\end{equation}
Note that $\langle.\rangle=:\langle\psi_{n}^{0}|\langle\psi_{n}^{cm}|.|\psi_{n}^{cm}\rangle|\psi_{n}^{0}\rangle$ and also
\begin{equation}
	H_{nz}^{cm}=\frac{{H_{n}^{cm}}}{e^{-z}}=\frac{p_{n}^{2}}{2me^{-z}}=\frac{\mathcal{P}_{nz}^{2}}{2m},
\end{equation}
where $(\mathcal{P}_{nz})^{2}=\frac{g_{ij}p_{n}^{i}p_{n}^{j}}{e^{-z}}$ with $(i, j =1,2,3)$

Therefore to derive time dilation and by noting the previous section considerations, one can write the average proper time read
by clock $A$ conditional on clock $B$ indicating the time $\tau_{B}$ in near horizon region supported by conformal metric $(\ref{tor})$ as follows
\begin{equation}\label{dilationzn}
	\langle T_{A}\rangle=\left(1-\frac{\langle e^{z}H_{Az}^{cm}\rangle-\langle e^{z}H_{Bz}^{cm}\rangle}{mc^{2}}\right)\tau_{B},
\end{equation}
Now suppose that the center-of-mass of both clocks are prepared in Gaussian state localized around an average momentum $\bar{\mathcal{P}}_{nz}$ with spread $\Delta _{n}>0$.
\begin{equation}
	|\psi_{n}^{cm}\rangle=\frac{1}{\pi^{1/4}\sqrt{\Delta_{n}}}\int d\mathcal{P}exp(\frac{-(\mathcal{P}-\bar{\mathcal{P}}_{nz})^{2}}{2\Delta_{n}^{2}})|\mathcal{P}_{nz}\rangle,
\end{equation}
which leads to
\begin{equation}\label{spread}
	\langle e^{z}H_{nz}^{cm}\rangle=e^{z}\frac{\bar{\mathcal{P}}_{nz}^{2}}{2m}+e^{z}\frac{\Delta_{n}^{2}}{4m}.
\end{equation}
By inserting equation $(\ref{spread})$ into $(\ref{dilationzn})$ the quantum average time
dilation between two clocks A and B in the near horizon region of a black hole described by metric $(\ref{tor})$ is obtained
\begin{equation}
	\langle T_{A}\rangle=\left[1-e^{z}\frac{\bar{\mathcal{P}}_{Az}^{2}-\bar{\mathcal{P}}_{Bz}^{2}+\frac{1}{2}(\Delta_{A}^{2}-\Delta_{B}^{2})}{2m^{2}c^{2}}\right]\tau_{B}.
\end{equation}
In the classical limit we have again the equation $(\ref{classicaldilation})$ and when one closes to the horizon of black hole $z \rightarrow -\infty$ we have $\tau_{A}=\tau_{B}$ which claims that there is no time dilation on the horizon.

\section{Conclusion}
In this paper, we have considered the Page and Wootters mechanism in the near-horizon region of the Schwarzschild black hole which is recognized by a static observer standing in the near horizon region. The observer considers the near horizon region like as Rindler observer. It is indicated that $N$ relativistic particles in this curved space-time can be considered by the PaW mechanism and an average quantum time dilation between two clocks $A$ and $B$ is obtained. This relation is derived for the Rindler background and also for the near horizon region of the black hole supported by conformal metric in tortoise coordinate. The quantum time dilation formula between two clocks is identified with well-known classical time dilation in general relativity for both cases.

 %%%%%%%%%%%%%%%%%%%%%%%%%%%%%%%%%%%%%%%%%%%%%%%%%%%%%%%%%%%%%%%%%%%%%%%%%%%%%%%%%%%%%%%%%%%%%%%%%%%%%%%%%%%%%%%%%%%%%%%%

\section*{Acknowledgements}

The paper is dedicated to the fond memory of one of the authors, Farhad Darabi who passed away on 13 Sept 2022.

This work is based upon research funded by Iran National Science Foundation
(INSF) under project No 99033073.

%%%%%%%%%%%%%%%%%%%%%%%%%%%%%%%%%%%%%%%%%%%%%%%%%%%%%%%%%%%%
%%%%%%%%%%%%%%%%%%%%%%%%%%%%%%%%%%%%%%%%%%%%%%%%%

\end{document}